\documentclass[journal,twoside,web]{IEEEtran}
\usepackage{cite}
\usepackage{amsmath,amssymb,amsfonts}
\usepackage{algorithmic}
\usepackage{romannum}
\usepackage{graphicx}
\usepackage{multirow}
\usepackage{pifont}
\usepackage{textcomp}
\makeatletter
\newcommand{\RNum}[1]{\uppercase\expandafter{\romannumeral #1\relax}}
\makeatother

\begin{document}
\title{PCDAL: A Perturbation Consistency-Driven Active Learning Approach for Medical Image Segmentation and Classification}
\author{Tao Wang, Xinlin Zhang, Yuanbo Zhou, Junlin Lan, Tao Tan, Min Du, Qinquan Gao and Tong Tong
\thanks{Tao Wang, Xinlin Zhang, Yuanbo Zhou, Junlin Lan, Min Du, Qinquan Gao and Tong Tong are with the college of physics and information engineering, University of Fuzhou University, Fuzhou 350108, China (e-mail: ortonwangtao@gmail.com; xinlin1219@gmail.com; webbozhou@gmail.com; smurfslan@qq.com; dm$\_$dj90@163.com; gqinquan@imperial-vision.com; ttraveltong@imperial-vision.com).}
\thanks{Tao Tan is with the Faculty of Applied Science, University of Macao Polytechnic University, Macao 999078 (e-mail: taotanjs@gmail.com).}
}

\maketitle

\begin{abstract}
In recent years, deep learning has become a breakthrough technique in assisting medical image diagnosis. Supervised learning using convolutional neural networks (CNN) provides state-of-the-art performance and has served as a benchmark for various medical image segmentation and classification. However, supervised learning deeply relies on large-scale annotated data, which is expensive, time-consuming, and even impractical to acquire in medical imaging applications.
Active Learning (AL) methods have been widely applied in natural image classification tasks to reduce annotation costs by selecting more valuable examples from the  unlabeled data pool. However, their application in medical image segmentation tasks is limited, and there is currently no effective and universal AL-based method specifically designed for 3D medical image segmentation.
To address this limitation, we propose an AL-based method that can be simultaneously applied to 2D medical image classification, segmentation, and 3D medical image segmentation tasks.
We extensively validated our proposed active learning method on three publicly available and challenging medical image datasets, Kvasir Dataset, COVID-19 Infection Segmentation Dataset, and BraTS2019 Dataset. The experimental results demonstrate that our PCDAL can achieve significantly improved performance with fewer annotations in 2D classification and segmentation and 3D segmentation tasks. The codes of this study are available at  https://github.com/ortonwang/PCDAL.
\end{abstract}


\section{Introduction}
\label{sec:introduction}
\IEEEPARstart{I}{n} the early stages, morphology-based methods such as clustering\cite{vasuda1713improved}, edge detection\cite{patil2013medical}, and threshold segmentation\cite{wang2015hybrid} were frequently utilized to process medical images for auxiliary diagnosis. However, with the advancement of artificial intelligence and deep learning, various technologies in computer vision have has been widely applied in medical image processing and analysis in recent years. This has made a significant impact on clinical practice, particularly in disease diagnosis and treatment planning, by enabling accurate identification of abnormal areas in medical images such as endoscopic images, X-ray images, and 3D Computed Tomography (CT) images\cite{wang2021review}. Accurate image segmentation in clinical medicine can provide clinicians with valuable reference information that enabling them to make diagnostic decisions quickly, accurately, and efficiently\cite{anwar2018medical}. Additionally, the objective information provided by computer methods can avoid human subjectivity\cite{sun2019deep}.

The Convolutional Neural Networks (CNN), particularly Fully Convolutional Networks, have been demonstrated to be effective \cite{wang2021pairwise} widely used for image segmentation and classification tasks. The ResNet\cite{He_2016_CVPR} network utilizes residual modules and greatly improves the accuracy of classification algorithms. U-Net\cite{ronneberger2015u} is a deep learning network with an encoder-decoder structure that has been widely applied in medical image segmentation. Building upon U-Net, Res-UNet\cite{xiao2018weighted} further improves the performance of U-Net by introducing residual modules in the encoder. To further expand the application of CNN in medical imaging data, Çiçek et al. proposed 3D U-Net\cite{cciccek20163d}, has shown excellent performance in 3D medical image segmentation tasks. With the further development of algorithms, the Vision Transformer (ViT)\cite{dosovitskiy2020image} was successfully applied to perform image recognition tasks and achieved good results. Based on ViT, Liu et al. developed Swin-Transformer\cite{liu2021swin} and further improved the performance. 

Most existing methods are supervised learning methods that require large-scale annotated data for training in order to achieve impressive performance. However, in medical imaging applications, it is often time-consuming, expensive, and even impractical to acquire such a huge annotated dataset\cite{jimenez2018capsule}. This is because, first, the acquisition of a large number of medical images is often time-consuming and expensive. Moreover, only sophisticated radiologists with years of experience are capable of annotating medical images\cite{kohli2017medical}.

Many studies have been performed to exploit the power of deep learning with limited annotated datasets for training.	
One notable approach is semi-supervised learning, which builds upon fully supervised learning and utilizes unlabeled data to enhance the performance of the models. In recent years, a lot of semi-supervised learning methods have been proposed, with notable examples including the works of SA-CPS\cite{9931157}, CAT\cite{10038734}, ASE-Net\cite{9966841} and SCANet\cite{9837087}. These approaches have partially alleviated the burden of data annotation. However, compared to supervised learning, semi-supervised learning often demands higher hardware requirements. Under equivalent batch size conditions, the semi-supervised learning frameworks typically consume 1.5 times or even more GPU memory for model training, resulting in longer training times. Consequently, these factors have somewhat hampered the widespread adoption of semi-supervised learning methods.

In contrast to semi-supervised learning, Active learning (AL) methods serve the annotation process by strategically selecting samples with maximum information and training utility \cite{tuia2009active}\cite{tuia2011survey}. It has been demonstrated that AL can significantly reduce the amount of annotated data required for image classification tasks \cite{gal2017deep}\cite{tran2019bayesian}\cite{sinha2019variational}. 
During the process of AL, individuals selectively choose data for annotation. After annotation is completed, the weights of the deep learning model obtained from annotated data are utilized to focus on the subsequent round of data annotation. This iterative process continues until satisfactory results are achieved.
Many efforts have been made to develop AL methods for image classification tasks. One commonly adopted approach is Maximum Entropy, which involves selecting samples that maximize the prediction entropy\cite{gal2017deep}. Additionally, diversity-based sampling\cite{patel2021hyperspectral} and kernel-based sampling\cite{shi2019active} have been employed as alternative methods. These techniques aim to select representative samples based on the underlying distribution of the data. Razvan Caramalau leveraged graph node embeddings and their confidence scores, adapting sampling techniques such as CoreSet and uncertainty-based methods to query the nodes\cite{Caramalau_2021_CVPR}. Amin Parvaneh et al. proposed a simple AL method based on the interpolation between labelled and unlabelled samples\cite{Parvaneh_2022_CVPR}.

In the context of 2D medical image classification and segmentation, specific methods have been developed. Samarth et al. proposed VAAL, which combines Variational Autoencoder (VAE) with adversarial networks to achieve an active selection of unlabeled data for image classification and segmentation\cite{sinha2019variational}. Meanwhile, Asim et al. introduced a novel sampling method, O-MedAL, which queries unlabeled examples that maximize the average distance to all training set examples\cite{smailagic2020medal}. Yang et al. utilized uncertainty information to enhance the performance of a retrained model, demonstrating excellent results in classification tasks\cite{yang2016active}. 	

Although those methods have yielded promising results, they fail to consider the inherent characteristics of medical images. For instance, medical images are commonly acquired from various imaging devices, such as CT scans, X-rays, endoscopies, etc., which requiring us to consider the generality of the methods. Furthermore, in the domain of AI-assisted medical image processing, image segmentation holds significant importance, particularly in the context of 3D medical image segmentation. However, currently, there is a dearth of AL-based methods that can be effectively applied to both 2D and 3D medical image segmentation tasks simultaneously.

Therefore, we propose A Perturbation Consistency-Driven Active Learning Approach (PCDAL), an AL-based method that is applicable to both 2D medical image classification and segmentation tasks as well as 3D medical image segmentation tasks. The objective is to identify more suitable data from the database for labeling, aiming to enhance algorithm performance.
In the implementation of this framework, we leverage the perturbation consistency of deep learning algorithm models to develop a perturbation consistency evaluation module that calculates perturbation errors. We sort unlabeled samples based on their respective perturbation error values and give priority to labeling data that is more significantly affected by perturbation.

We performed simulation experiments on three datasets: the Kvasir Dataset, the COVID-19 Infection Segmentation Dataset, and the BraTS2019 Dataset. The results from the segmentation and classification datasets demonstrate that using our proposed PCDAL framework to select more suitable data for annotation can lead to improved performance with fewer data annotations.
\section{Relate Works}

\subsection{Medical image classification and segmentation}
Early medical image classification methods primarily focused on machine learning\cite{erickson2017machine} and clustering\cite{vasuda1713improved} methods. However, these methods faced significant challenges when processing complex medical image data. Recently, deep learning-based methods have become increasingly popular, exhibiting excellent performance in medical image classification tasks such as skin lesion classification, endoscopic imaging, and lung CT image classification. Since the proposal of AlexNet\cite{krizhevsky2017imagenet} in 2012, more efficient and deep CNNs have been developed and applied to medical image classification tasks. ResNet, proposed by He et al., using stacked residual blocks, became a cornerstone of image classification tasks, and later optimized by researchers. DenseNet\cite{huang2017densely}, EfficientNet\cite{tan2019efficientnet}, and TransMed\cite{dai2021transmed}, which combined CNN with the Swin-Transformer, were successively proposed and applied to medical image classification tasks.

In early medical image segmentation tasks, the primary methods focused on using threshold segmentation\cite{emre2013lesion} and machine learning. Since the proposal of U-Net, it has become the benchmark of medical image segmentation tasks due to its outstanding performance and excellent generalizability, making it suitable for most medical image datasets. Res-UNet, proposed by Xiao et al., combined the U-Net with stacked residual blocks in the encoding part, while Zhou et al. further optimized U-Net's skip connection and proposed UNet++\cite{zhou2018unet++}. These improvements resulted in more accurate segmentation performance on some datasets.
Due to the excellent generalizability of the U-shaped architecture, 3D U-Net was proposed for 3D medical image segmentation. Currently, CNN-based U-shaped segmentation networks have achieved enormous success in medical image segmentation tasks. Various algorithms with different parameter sizes have been proposed such as SMU-Net\cite{9551285}, UNeXt\cite{valanarasu2022unext}, SpineParseNet\cite{9201093} and SegFormer\cite{xie2021segformer}. However, these optimization methods often increase the algorithm's parameter size and require higher hardware requirements. Furthermore, on small-scale datasets, the lack of annotated data leads to overfitting, resulting in unsatisfactory results. To improve the accuracy of intelligent medical diagnosis models, it is necessary to optimize both algorithm models and data. The AL can reduce the amount of labeled data to some extent, thus alleviating the burden of data labeling in the medical field.

\subsection{Active learning}

Active learning is a popular approach in machine learning that aims to improve the accuracy of models by selecting informative samples to label. One of the most commonly used sample selection strategies in active learning is uncertainty sampling, which selects samples that are difficult for the model to label.
Maximum Entropy\cite{wang2014new}\cite{gal2017deep} is a heuristic frequently used in this approach, aiming to selects samples that maximize predictive entropy. In addition to uncertainty sampling, there are other sample selection strategies, such as diversity-based sampling\cite{patel2021hyperspectral}, kernel-based sampling\cite{shi2019active}, and so on. These methods usually select representative samples based on the distribution of the data to improve the model's generalization ability.
Bayesian Active Learning by Disagreement (BALD)\cite{houlsby2011bayesian} and its batch version (i.e., BatchBALD)\cite{kirsch2019batchbald} selected samples that maximize the reduction in expected entropy \cite{pinsler2019bayesian} by computing the mutual information between model predictions and model parameters. 
Wei et al. proposed a good acquisition function for the uncertainty component related to the model's expected correctness score, based on the BEMP\cite{tan2021diversity} method, and demonstrated excellent performance in active learning in text classification tasks using CoreMSE and CoreLog.
Razvan Caramalau et al. introduced CoreGCN and UncertainGCN constructed on a sequential Graph Convolution Network (GCN) and utilized the graph node embeddings and their confidence scores and adapt sampling techniques such as CoreSet and uncertainty-based methods to query the nodes\cite{Caramalau_2021_CVPR}. Amin Parvaneh et al. proposed ALFA-Mix. Unlabelled instances were identified with sufficiently-distinct features by seeking inconsistencies in predictions resulting from interventions on their representations\cite{Parvaneh_2022_CVPR}.

Ensemble methods, such as Monte Carlo dropout\cite{gal2016dropout} and deep ensembles\cite{lakshminarayanan2017simple}, have successfully been used to obtain better uncertainty estimates with deep learning. More sophisticated techniques are being developed, such as MCMC, hybrid, and deterministic approaches\cite{wilson2020bayesian}\cite{ashukha2020pitfalls}. However, plain ensembling remains a competitive and simple method for deep learning\cite{ashukha2020pitfalls}. The Variational Adversarial Active Learning (VAAL)\cite{sinha2019variational} proposed a pool-based semi-supervised active learning method, which uses VAE and adversarial networks to learn the latent space, so that the adversarial network learns how to distinguish differences in the latent space when VAE tries to deceive it by predicting that all data points come from the labeled pool, achieving the effect of active learning. This method has been applied to segmentation and classification tasks in 2D image analysis, but it has not been employed for 3D image segmentation tasks.
MedAL\cite{smailagic2018medal} achieved the effect of active learning by querying unlabelled examples to maximize the average distance between all training set samples in the learned feature space. O-MedAL\cite{smailagic2020medal} extended this method by prioritizing unlabelled examples that have the largest distance to the centroid of all training set samples in the learned feature space for data annotation.
Overall, active learning has shown promise in reducing the amount of labeled data needed to train machine learning models, especially in cases where the annotated data is limited. Different sample selection strategies and ensemble methods have been proposed to improve the performance of active learning in various domains. Further research is needed to develop more efficient and effective active learning methods for machine learning applications.

\begin{figure}[h]
	\centering
	\includegraphics[scale=0.38]{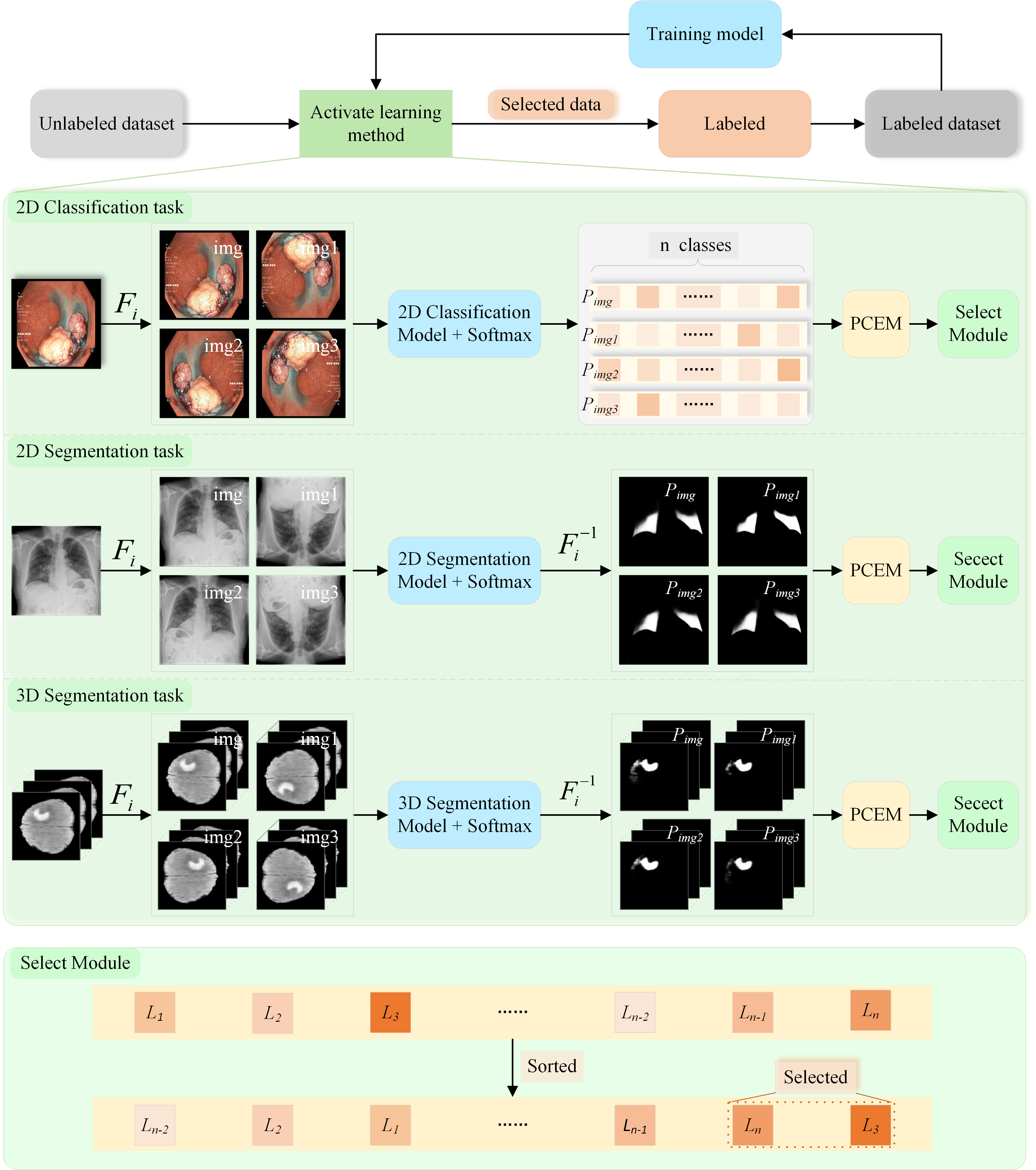}		
	\caption{PCDAL for active learning of medical image. The $F$ refers to the transformation-consistent flip processing,including horizontal flipping, vertical flipping and the combination of horizontal and vertical flipping. The $P$ represents the confidence scores generated by the network model for each class or pixel in the classification or segmentation task. The PCEM means Perturbation Consistency Evaluation Module, $L$ represents the perturbation error value, with the depth of the background color increasing proportionally to the magnitude of the error.}
\end{figure}

\section{The Proposed Method}
\subsection{Overall architecture design}
Figure 1 illustrates the overall architecture of the PCDAL framework proposed in our study, which is an active learning approach suitable for 2D medical image classification and segmentation, as well as 3D medical image segmentation tasks.
Specifically, first, the proposed approach involves initially training a CNN on a limited set of annotated data.
Once the training is complete, the existing model weights are used to infer both the unlabeled and augmented data, thereby generating multiple predictions for the same image.
We develop the following strategies to select reliable annotated images for training. We normalize the predicted results with the softmax operation to ensure that the predictions are comparable and to reduce the influence of outliers. Moreover, we introduce a Perturbation Consistency Evaluation Module (PCEM), which assesses the consistency among multiple predictions generated for the same image. Last, we prioritize the unlabeled data set based on perturbation consistency. Only samples with high perturbation impact are selected for annotation. The annotated images are then merged with the original small annotated data set. We iterate this process to expand the training dataset and improve the performance of the model with fewer annotations. This iterative process can be repeated until the accuracy of the approach meets the anticipated requirements, and subsequently, the iteration can be terminated.\\
\subsection{Flip-based data perturbation method}
The proposed framework, PCDAL, is based on the consistency of perturbation results obtained from the algorithmic model on data to select meaningful data for annotation. To achieve this, we employed the widely used flip-based perturbation method, where each image is subject to horizontal flipping, vertical flipping and the combination of horizontal and vertical flipping. These operations are easy to implement and have low hardware requirements. Furthermore, the experimental results demonstrate outstanding performance. Importantly, these perturbation operations do not alter the semantic information contained in the images, rendering them suitable for various 2D or 3D medical image. The specific effect of perturbation of image can be found in Figure 1. Unlike traditional data augmentation methods that aim to improve the model's performance, this study focuses on augmenting unlabeled data to select valuable data for annotation. Figure 2 provides a clearer visualization of the impact of perturbations on the predictions for images. The effects of perturbations on the algorithm's predictions are more intuitively demonstrated in the figure. 
It can be observed that under the same weights, the perturbation applied to the image leads to certain variations in the predicted values, particularly in the area indicated by the red arrow. The magnitude of prediction deviation caused by perturbations indicates the model's lower confidence in these data samples, implying a relative scarcity of such data within the training set. This underscores the importance of prioritizing the annotation of these data.
Compared to the data with less influence from perturbations, it is assumed that assigning a label to those with more influence from perturbations holds greater value. The experimental results presented later in our paper further validate this observation.
\begin{figure}[!h]
	\centering
	\includegraphics[scale=1.7]{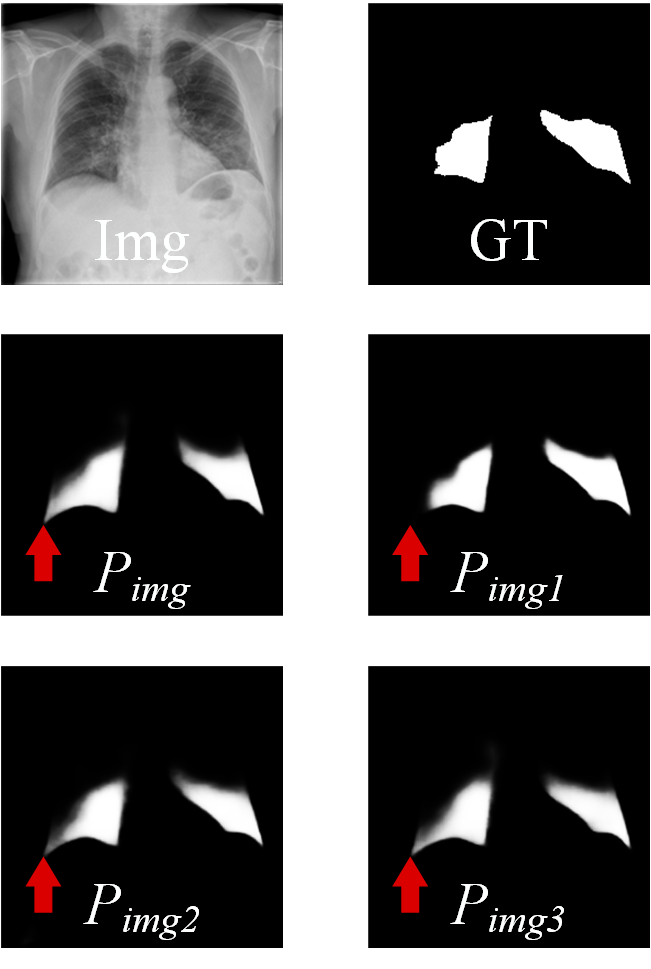}
	\caption{Differential predicted result of the images after perturbations.}
\end{figure}
\subsection{Perturbation Consistency Evaluation Module}
To emphasize the impact of perturbations on the consistency of prediction results, we adopt an error calculation method based on the Mean Square Error (MSE) function, as illustrated in Equation (1):
\begin{equation}
P_{average}=\frac{1}{4} \sum_{i=1}^{4}P_{img_{i}} ,			
\end{equation}
\begin{equation}
Pturb\ cls=\frac{1}{4} \sum_{i=1}^{4}(P_{img_{i}} - P_{average})^{2} 
\end{equation}
\begin{equation}
Pturb\ seg=\frac{1}{N} \sum_{j=1}^{N} Pturb\ cls_{j}
\end{equation}
where $P_{img_{i}}$ means the prediction results of the input image processes by the model, $N$ is the number of pixels in the image, $Pturb$ $cls$ and $Pturb$ $seg$ means perturbation errors of the classification task and segmentation task, respectively.
Firstly, we compute the average of the prediction results for the input image and the three perturbed images, denoted as $P_{average}$. In the classification task, we calculate the error between each result and $P_{average}$ and square the errors to emphasize the data that were more significantly affected by perturbations. Finally, we obtained the average of the four squared errors. 
In the segmentation task, we performed the same operation for all $N$ pixels to obtain the average error of $N$ pixels. This method enables us to effectively evaluate the consistency of the perturbation results and identify the most valuable data for annotation.

\subsection{Application in classification Tasks}
For n-classification tasks, the network processes each image resulting in a 1$\times$n tensor. Although the perturbations used to augment the images may affect the confidence levels of each category in n-classification tasks, they do not affect the distribution of confidence levels. Therefore, in classification tasks, we use the PCEM to calculate the consistency between predicted results for input data and perturbed data. After inferring all unlabeled images, we obtain the consistency error value for each image. Finally, we ranked all images based on their error values and prioritized the annotation of images with high perturbation impact (HPI).

\subsection{Application in segmentation tasks}
For 2D/3D segmentation tasks, the input images that have been fliped are in the same fliped state after being processed by the network. Thus, the first step is to perform an inverse transformation that is equivalent to the flip on the images to maintain their original orientation. Afterward, the PCEM calculates the consistency between the segmentation results of the perturbed data and the original data, which yields the perturbation error value for each image. Similar to classification tasks, unannotated data is ranked based on their error values, and annotation is prioritized for data with HPI.
\section{Experiments}	
Three diverse biomedical image datasets acquired from different devices were used to evaluate the performance of our PCDAL framework for both segmentation and classification tasks. By utilizing multiple datasets, we sought to assess the generalizability of the proposed approach and validate its effectiveness across varied biomedical imaging applications. The experimental results obtained through this evaluation provide a comprehensive understanding of the capabilities and limitations of our proposed method, thereby contributing to the advancement of medical image analysis research.
\subsection{Datasets}
\textbf{Kvasir dataset}: The Kvasir Dataset\cite{pogorelov2017kvasir}, used in our study, was originally released as part of the medical multimedia challenge presented by MediaEval. The images in the dataset were captured via an endoscopy equipment from inside the gastrointestinal tract and were annotated and verified by professional doctors. The dataset comprises eight classes, each containing 1000 images, based on three anatomical landmarks (z-line, pylorus, cecum), three pathological findings (esophagitis, polyps, ulcerative colitis), and two other classes (dyed and lifted polyps, dyed resection margins) related to the polyp removal process. To ensure balance across categories, we randomly divided the dataset of 4000 images into a validation set of 2000 images and a test set of 2000 images. The images in the dataset vary in size from 576×576 to 1072×1920 pixels.

\textbf{COVID-19 Infection Segmentation Dataset}: The dataset is a subset of COVID-QU-Ex\cite{chowdhury2020can}\cite{degerli2021covid}\cite{tahir2021covid}\cite{rahman2021exploring} collected by researchers from Qatar University. It contains a total of 2913 CXRs with corresponding lung masks from COVID-QU-Ex dataset and corresponding infection masks from QaTaCov19 dataset, in addition to 1456 normal and 1457 non-COVID-19 chest X-rays. The labeling of this dataset has been carried out by professional physicians to ensure the accuracy and reliability of the dataset.
The resolution of all images in this dataset has been fixed to 256$\times$256. Specifically, the dataset consists of 3728 images in the training set, 932 images in the validation set, and 1166 images in the test set. It is noteworthy that the infection part of the data is used as positive examples, while the remaining images are used as background for segmentation purposes.

\textbf{BraTS2019 Dataset}: The BraTS2019\cite{menze2014multimodal} dataset, which comprises brain MR scans of 335 glioma patients, is a publicly available resource. This dataset comprises MRI scan images from different clinical centers, including T1, T2, T1 contrast enhanced, and FLAIR sequences, along with corresponding tumor segmentations annotated by expert radiologists. In this study, we utilized the FLAIR modality to perform segmentation on the dataset. To partition the dataset, we employed a random split of 250, 25, and 60 scans for training, validation, and testing, respectively, in accordance with SSL4MIS\cite{ssl4mis2020}. In terms of pre-processing, we first cropped the zero-intensity region and subsequently rescaled the intensity of each scan to the range [0, 1].
\subsection{Implementation Details}
Our method was implemented using the Pytorch deep learning framework in Python. For the kvasir dataset, we used a ResNet34 backbone for classification, with an input image size of 512$\times$512.
For the COVID-19 dataset, we used a Res34-UNet backbone for segmentation, with an original image size of 256$\times$256. To increase the size of the training set and prevent overfitting, we used random flips and rotations as data augmentation techniques for both datasets. The encoders of both the classification and segmentation models were pre-trained on ImageNet\cite{deng2009imagenet}.
For the BraTS2019 dataset, we utilized a 3D U-Net as the backbone segmentation network. The model was designed to take randomly cropped patches as input, with a patch size of 96$\times$96$\times$96. Random cropping, flipping, and rotation were employed as data augmentation techniques to enhance the generalization ability of both models.

We trained our model on a Nvidia RTX 3090 GPU with 24GB of memory. For the Kvasir and COVID-19 Infection Segmentation Dataset, the batch size was set to 16 and the popular AdamW optimizer with a learning rate of 1e-4 for the backpropagation of the models and we utilized the cross-entropy loss. While for the BraTS2019 dataset, the batch size was set to 4 and we adopt the SGD optimizer with a learning rate of 1e-1. To account for class imbalance, we utilized a weighted Dice loss with a ratio of 1:1, in addition to the cross-entropy loss. Additionally, we employed a CosineAnnealing learning rate schedule for learning rate decay.

As our study employed a validation approach that involves randomly selecting a portion of the data to simulate a small annotated dataset, as in real-world applications, the method of randomly selecting a portion of the data can introduce randomness that may affect the experimental results due to random errors. To minimize such random errors and provide a more objective comparison of the effectiveness of different methods, a stratified five-fold cross-validation method was used to randomly select the training set data. During the experiment, 10$\%$ of the training set data was randomly selected as the initial dataset. Subsequently, 10$\%$ of data was added each time to supplement the data for iteration. By using the stratified five-fold cross-validation method, we aimed to provide a more rigorous and objective approach to compare the effectiveness of different methods and mitigate the impact of random errors introduced by the initial selection of data.

\subsection{Evaluation Metrics}
To evaluate the effectiveness of our approach in various biomedical imaging tasks, we employed the Precision (Pre) and Accuracy (Acc) as evaluation metrics for classification tasks. For 2D image segmentation tasks, we used the Dice coefficient (PA) and Pixel Accuracy (PA) to measure the effectiveness of segmentation. While for 3D image segmentation tasks,we used the the Dice coefficient and the 95$\%$ Hausdorff Distance (HD$_{95}$) to evaluate our method. The calculation formulas for these evaluation metrics were as follows:
\begin{equation}
Pre = \frac{TP}{ TP + FP},
\end{equation}
\begin{equation}
Acc = \frac{TP + TF}{TP + TN + FP + FN},
\end{equation}
\begin{equation}
Dice = \frac{2\times{TP}}{ 2\times{TP} + FN + FP},
\end{equation}
\begin{equation}
PA = \frac{TP + TN}{TP + TN + FP + FN},
\end{equation}
where TP is the number of true positives, TN is the number of true negatives, FP is the number of false positives, and FN is the number of false negatives.
The HD$_{95}$ calculates the maximum distance between the contours of the ground truth and predicted results, which can be formulated as follows:
\begin{equation}
HD_{95} = max(h(A,B),h(B,A),
\end{equation}
\begin{equation}
h(A,B) =  \max_{a \in A} \left\{{{\min_{b \in B}\left\{\Vert a-b \Vert\right\}}}\right\},
\end{equation}
\begin{equation}
h(B,A) =  \max_{b \in B} \left\{{{\min_{a \in A}\left\{\Vert b-a \Vert\right\}}}\right\}
\end{equation}
where A and B denote the contours of the ground truth and predicted results, respectively, and $h$(A,B) denotes the unidirectional Hausdorff distance from A to B.

\subsection{Experiment results on the Kvasir}
We compared our PCDAL with the state-of-the-art active learning methods. Figure 3 demonstrates the superiority of our method. Table \RNum{1} shows the mean and standard deviation obtained by different methods and it reveals that active learning methods are more accurate and efficient than the random sample methods. During the experiment, a random sample of 10$\%$ of the training set data was used as the initial dataset for all active learning methods. As a result, the results of these methods were identical when 10$\%$ of the data was annotated. Methods based on Max Entropy, CoreLog, and other active learning techniques can improve the algorithm's performance to some extent. However, the accuracies of these methods on this dataset is limited compared to the accuracy achieved by randomly selecting data. Compared with these methods, our proposed PCDAL approach has significant advantages. As shown in Table \RNum{1}, our PCDAL method achieved an accuracy of 93.06 by selecting 30$\%$ of the data as the training set, which is better than randomly selecting 50$\%$ of the data. Our approach not only saves 20$\%$ of data labeling efforts but also improves the algorithm's accuracy. Moreover, these results indicates that constructing a dataset by selecting low perturbation impact (LPI) data has limited implications on algorithmic improvement. This suggests that our proposed method is capable of effectively identifying relatively unimportant samples in medical image classification tasks. In practical applications, excluding these samples can also mitigate the waste of labeling resources to some extent. Meanwhile, with the increase of data annotated, the improvement in accuracy gradually diminishes. When 50$\%$ of the data is annotated, the accuracy does not show significant improvement compared to that with 40$\%$ of the data is annotated. This also demonstrates the strong ability of our PCDAL in selecting informative samples. After several data selections, the proportion of remaining effective data gradually decreases.
Furthermore, our stratified five-fold cross-validation results based on our PCDAL method exhibit the lowest standard deviation, demonstrating the stability and effectiveness of our method while ensuring its effectiveness. Therefore, our PCDAL method can provide a more rigorous and objective approach to compare the effectiveness of different methods, and to mitigate the impact of random errors introduced by the initial selection of data.\\

\begin{table}[h]
	\centering
	\caption{Experimental recults of different active learning methods on the Kvasir dataset under different rates of labeled images.}
	\scalebox{0.7}{
	\begin{tabular}{c|c|ccccc}
		\hline
		Method                                               & Metric                & 10\%                         & 20\%                & 30\%                & 40\%                & 50\%                \\ \hline
		Our-LPI                                              & \multirow{10}{*}{Acc$\uparrow$} & \multirow{10}{*}{89.40±0.94} & 89.83±0.53          & 90.22±0.59          & 90.79±0.43          & 91.29±0.53          \\
		VAAL\cite{sinha2019variational}     &                       &                              & 91.23±0.39          & 91.78±0.38          & 92.34±0.49          & 92.73±0.42          \\
		O-MedAL\cite{smailagic2020medal}    &                       &                              & 91.19±0.23          & 91.83±0.32          & 91.94±0.59          & 92.58±0.33          \\
		Randn                                                &                       &                              & 90.99±0.68          & 92.01±0.51          & 92.51±0.27          & 92.85±0.66          \\
		CoreMSE\cite{tan2021diversity}      &                       &                              & 91.32±0.68          & 92.11±0.43          & 92.33±0.35          & 92.37±0.49          \\
		CoreLog\cite{tan2021diversity}      &                       &                              & 90.93±0.52          & 92.16±0.30          & 92.82±0.48          & 92.31±0.53          \\
		CoreGCN\cite{Caramalau_2021_CVPR} &                       &                              & 90.89±0.49          & 92.31±0.54          & 92.96±0.28          & 93.17±0.20          \\
		Max-Entropy\cite{gal2017deep}       &                       &                              & 92.01±0.42          & 92.78±0.34          & 93.41±0.28          & 93.69±0.14          \\
		ALFA-Mix\cite{Parvaneh_2022_CVPR} &                       &                              & 91.58±0.49          & 92.81±0.59          & 93.18±0.66          & 93.23±0.48          \\
		Ours                                                 &                       &                              & \textbf{92.10±0.35} & \textbf{93.06±0.09} & \textbf{93.76±0.20} & \textbf{93.76±0.12} \\ \hline
		Our-LPI                                              & \multirow{10}{*}{Pre$\uparrow$} & \multirow{10}{*}{89.65±0.78} & 89.92±0.46          & 90.36±0.54          & 90.87±0.39          & 91.47±0.48          \\
		VAAL\cite{sinha2019variational}     &                       &                              & 91.35±0.41          & 91.98±0.25          & 92.47±0.46          & 92.80±0.42          \\
		O-MedAL\cite{smailagic2020medal}    &                       &                              & 92.02±0.36          & 92.17±0.46          & 92.74±0.35          & 93.18±0.32          \\
		Randn                                                &                       &                              & 91.32±0.67          & 92.48±0.53          & 92.56±0.27          & 92.91±0.66          \\
		CoreMSE\cite{tan2021diversity}      &                       &                              & 91.41±0.67          & 92.27±0.44          & 92.41±0.33          & 92.44±0.42          \\
		CoreLog\cite{tan2021diversity}      &                       &                              & 91.05±0.52          & 92.27±0.34          & 92.88±0.49          & 92.44±0.53          \\
		CoreGCN\cite{Caramalau_2021_CVPR} &                       &                              & 91.10±0.53          & 92.40±0.49          & 93.01±0.29          & 93.24±0.19          \\
		Max-Entropy\cite{gal2017deep}       &                       &                              & 92.17±0.36          & 92.82±0.33          & 93.44±0.27          & 93.77±0.13          \\
		ALFA-Mix\cite{Parvaneh_2022_CVPR} &                       &                              & 91.67±0.48          & 92.87±0.60          & 93.23±0.61          & 93.30±0.42          \\
		Ours                                                 &                       &                              & \textbf{92.16±0.33} & \textbf{93.18±0.11} & \textbf{93.83±0.17} & \textbf{93.82±0.11} \\ \hline
	\end{tabular}}
\end{table}

\begin{figure}[h]
	\centering
	\includegraphics[scale=0.5]{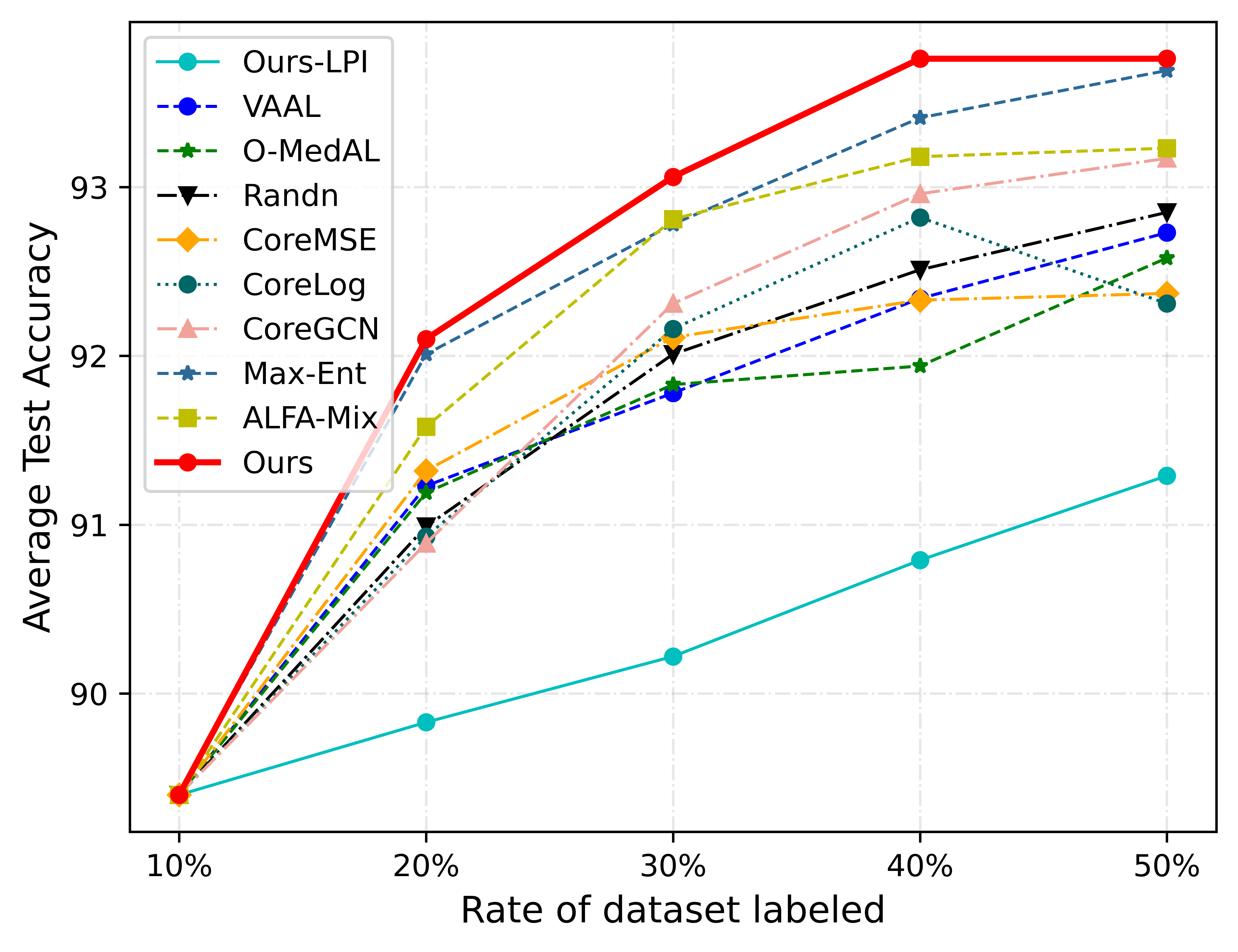}
	\caption{Classification accuracy on the Kvasir dataset.}
\end{figure}
\begin{table}[h]
	\centering
	\caption{Experimental recults on the COVID-19 Infection Segmentation data under different rates of labeled images.}
	\scalebox{0.75}{
	\begin{tabular}{c|c|ccccc}
		\hline
		Method & Metric & 10\%       & 20\%       & 30\%       & 40\%       & 50\%       \\ \hline
		Ours-LPI    & \multirow{4}{*}{Dice$\uparrow$}       &\multirow{4}{*}{89.26±1.19} & 91.03±0.95 & 91.23±0.77 & 92.55±0.17 & 92.84±0.11 \\
		Randn  &    &  & 91.36±1.09 & 92.84±0.67 & 93.38±0.18 & 93.45±0.2  \\
		VAAL\cite{sinha2019variational}   &        &  & 91.42±1.21 & 93.18±0.19 & 93.48±0.19 & 93.61±0.13 \\
		Ours   &        &  & \textbf{92.23±0.49} & \textbf{93.29±0.22} & \textbf{93.67±0.22} & \textbf{93.76±0.17} \\ \hline
		Ours-LPI    &\multirow{4}{*}{PA$\uparrow$}        & \multirow{4}{*}{97.15±0.16} & 97.43±0.11 & 97.42±0.04 & 97.56±0.04 & 97.57±0.03 \\
		Randn  &      & & 97.60±0.05 & 97.79±0.02 & 97.85±0.04 & 97.92±0.09 \\
		VAAL\cite{sinha2019variational}   &        &  & 97.56±0.06 & 97.78±0.02 & 97.86±0.03 & 97.92±0.01 \\
		Ours   &        & & \textbf{97.74±0.03} & \textbf{97.90±0.02} & \textbf{97.99±0.04} & \textbf{98.03±0.02} \\ \hline
	\end{tabular}}
\end{table}
\begin{figure}[h]
	\centering
	\includegraphics[scale=0.5]{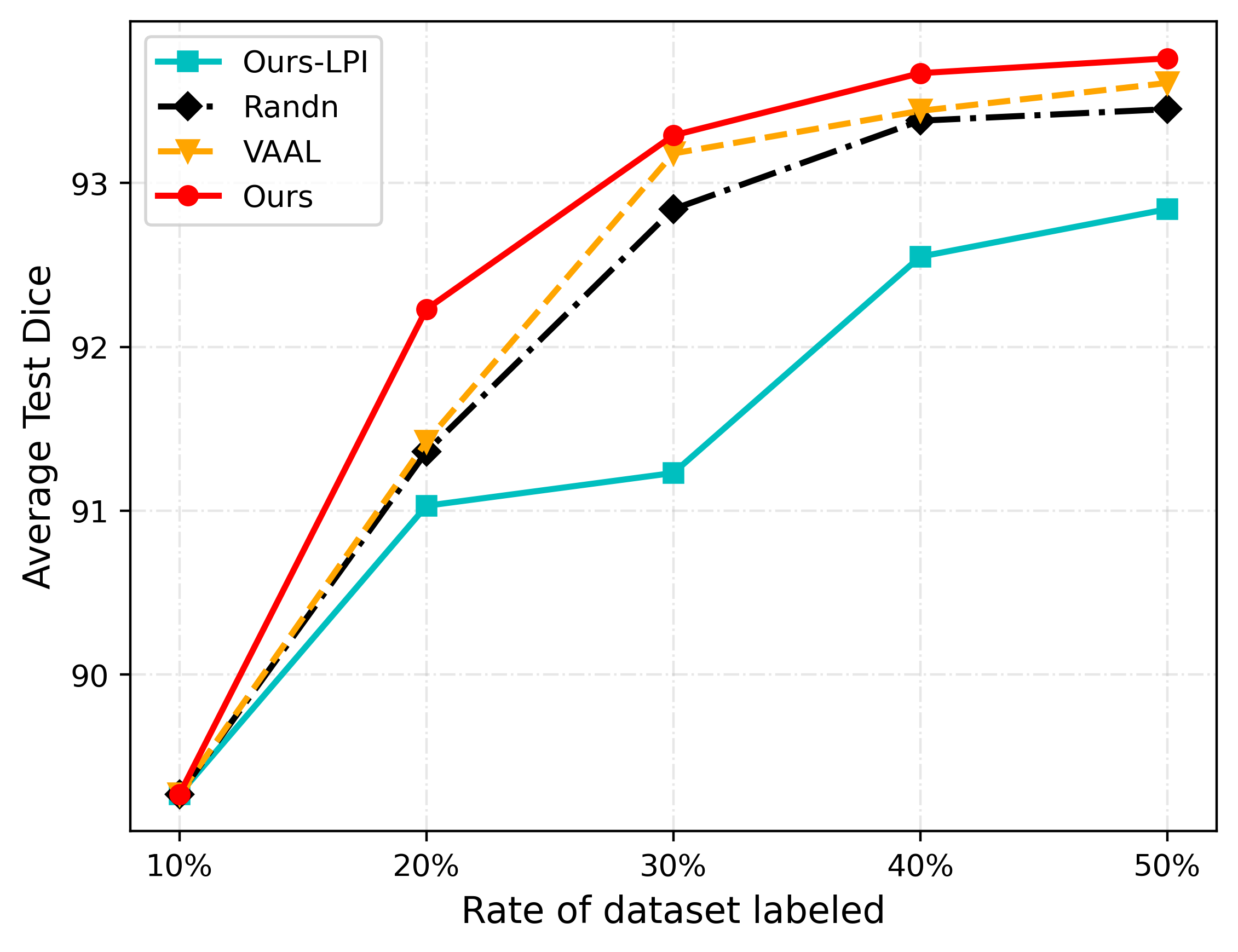}
	\caption{The Dice score for COVID-19 infection segmentation results. }
\end{figure}
\subsection{Experiment results on the COVID-19 Infection Segmentation}
From Figure 4, it is evident that our proposed PCDAL method significantly outperforms randomly selected annotated data in terms of Dice, given the same amount of annotated data. Table \RNum{2} shows the mean and standard deviation obtained by different methods. The results indicate that selecting data with HPI for annotation  results in significantly lower variance in the test dataset than randomly sampled data when the amout of labeled data are low. On the other hand, selecting LPI data has limited implications on algorithmic improvement. 
When 50$\%$ of the complete dataset was used for annotation, the LPI method achieved a Dice score of 92.84$\%$ and a PA score of 97.57$\%$. Surprisingly, these results were even lower than those obtained using the HPI method, which selected only 30$\%$ of the data for training but achieved a Dice score of 93.29$\%$ and a PA score of 97.9$\%$. 


Figure 5 depicts the segmentation results of the tested data after training using different strategies for data selection and annotation when 30$\%$ of the data is labeled. From the figure, it can be observed that training with data obtained by the PCDAL method produces better segmentation results than random selection. The effect of using VAAL will be slightly more accurate than that of randomly selecting data, but the effect is still relatively limited.
Conversely, the model trained by the LPI method results in more noise compared to the HPI method and the method of randomly selecting data in the image segmentation results. This indicates that our proposed PCDAL method can effectively distinguish between relatively effective and ineffective samples in 2D medical image segmentation tasks.\\
\begin{figure}[!h]
	\centering
	\includegraphics[scale=0.6]{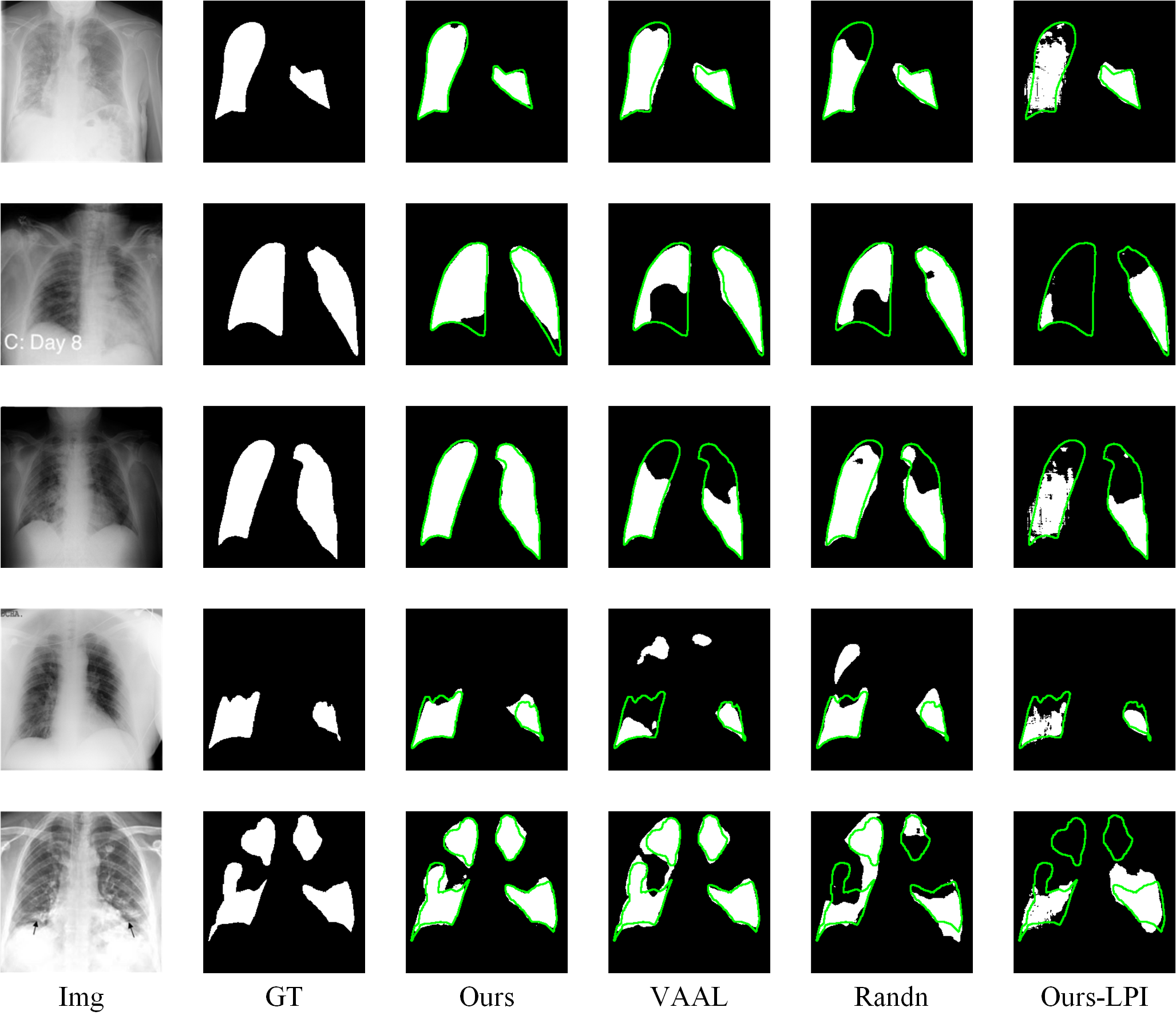}
	\caption{Comparision of different methods for data selection and annotation on the COVID-19 Infection Segmentation Data when 30$\%$ of the data is labeled. The edge contour of the Ground Truth is represented by the green line.}
\end{figure}
\subsection{Experiment results on the BraTS2019}
We further evaluated the application of our method in 3D segmentation tasks. The segmentation results on the BraTS2019 dataset are present in Figure 6 and Table \RNum{3}. From the figures, it is evident that our proposed PCDAL method exhibits excellent performance. With the same amount of labeled data, our method achieved the highest Dice score and the lowest HD$_{95}$ value.
Additionally, our PCDAL method achieved a Dice coefficient of 85.98 and a HD$_{95}$ of 9.38 when 30$\%$ of the data was selected as the training set. Notably, this performance is nearly indistinguishable from randomly selecting 50$\%$ of the data. As a result, our approach effectively reduces the annotation burden by approximately 20$\%$ in this regard.
In contrast, selecting LPI data for annotation resulted in limited improvement in performance which get the lowest Dice and highest HD$_{95}$. Figure 7 illustrates the segmentation results on the test dataset when 30$\%$ of the data is labeled using random selection, our PCDAL, and the LPI method with data selection. As observed from the figure, our proposed method achieves more precise segmentation results compared to the models generated by random and LPI methods. These models face issues in fully identifying lesions with subtle color differences. However, our method can overcome this problem to a certain extent. The significant difference in algorithm performance improvement between supplementing HPI data and LPI data further indicates the effectiveness of our PCDAL framework in distinguishing between valid and invalid data, thereby enhancing the algorithm's performance in medical image segmentation tasks.
\begin{figure}[!h]
	\centering
	\includegraphics[scale=0.6]{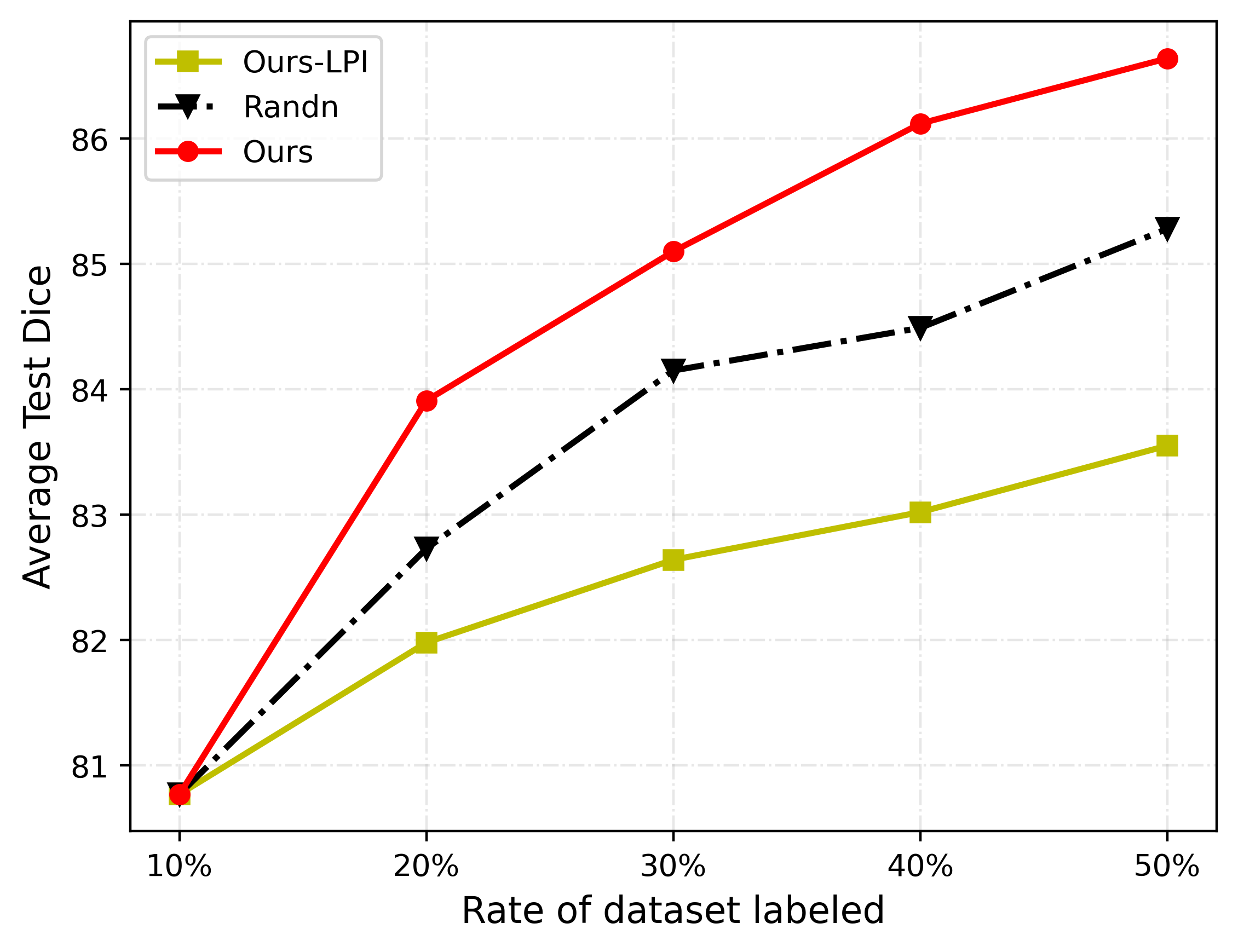}
	\caption{The Dice score for BraTS2019 segmentation results.}
\end{figure}
\begin{table}[h]
	\centering
	\caption{Experimental recults on the BraTS2019 under different rates of labeled images.}
	\scalebox{0.75}{
		\begin{tabular}{c|c|ccccc}
			\hline
			Method   & Metric                & 10\%                        & 20\%       & 30\%       & 40\%       & 50\%       \\ \hline
			Ours-LPI & \multirow{3}{*}{Dice$\uparrow$} & \multirow{3}{*}{80.77±0.94} & 81.98±0.41 & 82.64±0.68 & 83.02±0.93 & 83.55±0.61 \\
			Randn    &                       &                             & 82.73±0.59 & 84.15±0.41 & 84.49±0.82 & 85.28±0.61 \\
			Ours     &                       &                             & 83.91±0.61 & 85.10±0.55 & 86.12±0.45 & 86.64±0.51 \\ \hline
			LPI      & \multirow{3}{*}{HD$_{95}$$\downarrow$}   & \multirow{3}{*}{17.84±4.28} & 16.82±7.54 & 14.53±2.74 & 13.42±2.03 & 11.30±1.41 \\
			Randn    &                       &                             & 17.72±5.76 & 10.24±0.93 & 11.04±3.66 & 9.26±0.86  \\
			Ours     &                       &                             & 13.25±2.33 & 9.38±0.85  & 7.95±0.41  & 8.48±1.21  \\ \hline
	\end{tabular}}
\end{table}

\begin{figure}[h]
	\centering
	\includegraphics[scale=0.7]{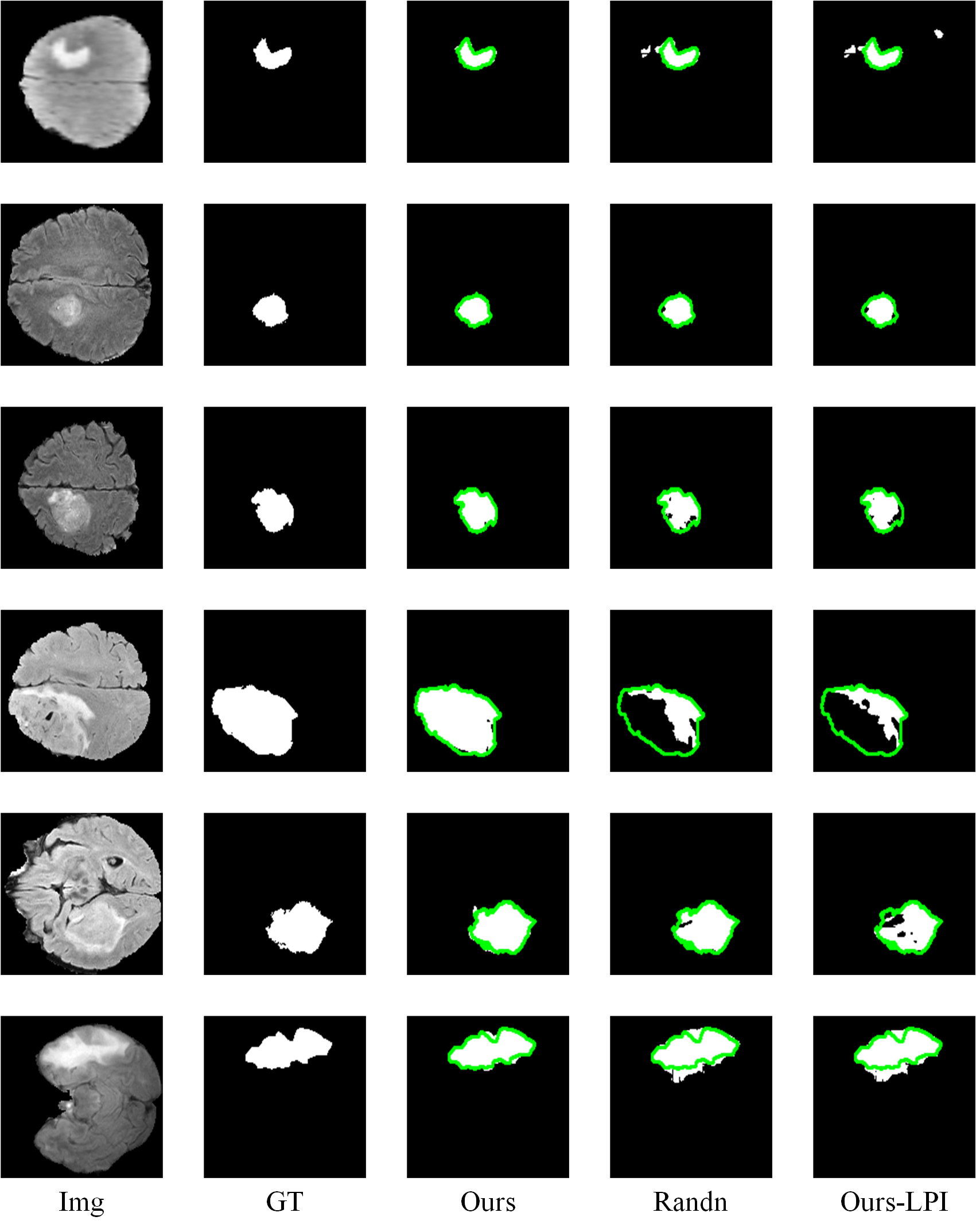}
	\caption{Comparision of different methods for data selection and annotation on the BraTS2019 dataset when 30$\%$ of the data is labeled. The edge contour of the Ground Truth is represented by the green line.}
\end{figure}


\subsection{Abulation stydy}
To explore the effects of different methods on the efficacy of our proposed PCDAL approach, we further performed an ablation study on the Kvasir dataset. Specifically, we investigated the impact of different perturbation methods on the data and the function used to calculate the perturbation error. 

\subsubsection{Effect of perturbation methods} While various methods exist for perturbing data, excessive perturbations may consume more computational resources without improving results. Careful selection and application of perturbation methods are essential for optimizing efficacy and efficiency in various applications, such as medical image analysis. Therefore, we compared the effects of different perturbation methods. We combined five different perturbation methods: horizontal flipping, vertical flipping, the combination of horizontal and vertical flipping, 90-degree clockwise rotation, and 180-degree clockwise rotation. Results presented in Table \RNum{4} show that the simultaneous adoption of horizontal flipping, vertical flipping, the combination of horizontal and vertical flipping can achieve the desired perturbation effect while slightly increasing the computational cost. 
Simultaneously employing horizontal flipping and vertical flipping also yielded promising performance, but they exhibited higher standard deviation compared to the combination of three flip.
On the other hand, further increasing the perturbations beyond this point did not significantly improve the results but increased the computational burden of the algorithm. Consequently, we selected the perturbation method that applies horizontal flipping, vertical flipping, and the combination of horizontal and vertical flipping to the input image.
\begin{table}[h]
	
	\centering
	\caption{Abulation study on the perturbation methods. F. V represents horizontal flipping of images, F. H means vertical flipping of images, F. V $\&$ F. H denotes the combination of both operations. R represents clockwise rotation of 90$^{\circ}$.}
	\scalebox{0.55}{
	\begin{tabular}{ccccc|c|ccccc}
		\hline
		\multicolumn{1}{c|}{F. V} & \multicolumn{1}{c|}{F. H} & \multicolumn{1}{c|}{F. V \& F. H} & \multicolumn{1}{c|}{R \& F. V} & R \& F. H & \multicolumn{1}{c|}{Metric} & \multicolumn{1}{c}{10\%}    & 20\%                & 30\%                & 40\%                & 50\%                \\ \hline
		\checkmark                           &                             &                                &                                     &                & \multirow{5}{*}{Acc$\uparrow$}        & \multirow{5}{*}{89.40±0.94} & 91.61±0.77          & 92.79±0.36          & 93.19±0.52          & 93.69±0.31          \\
		\checkmark                           & \checkmark                           &                                &                                     &                &                             &                             & 91.72±0.61          & \textbf{93.16±0.25} & 93.45±0.28          & \textbf{94.04±0.13} \\
		\checkmark                           & \checkmark                          & \checkmark                              & \checkmark                                  &                &                             &                             & 91.75±0.26          & 93.10±0.60          & 93.59±0.16          & 93.84±0.24          \\
		\checkmark                           & \checkmark                           & \checkmark                             & \checkmark                                   & \checkmark              &                             &                             & 91.68±0.44          & 93.05±0.23          & 93.35±0.42          & 93.48±0.25          \\
		\checkmark                           & \checkmark                          & \checkmark                              &                                     &                &                             &                             & \textbf{92.10±0.35} & 93.06±0.09          & \textbf{93.76±0.20} & 93.76±0.12          \\ \hline
		\checkmark                           &                             &                                &                                     &                & \multirow{5}{*}{Pre$\uparrow$}        & \multirow{5}{*}{89.65±0.78} & 91.72±0.69          & 92.83±0.33          & 93.29±0.44          & 93.73±0.31          \\
		\checkmark                           & \checkmark                           &                                &                                     &                &                             &                             & 91.82±0.62          & \textbf{93.20±0.23} & 93.51±0.27          & \textbf{94.07±0.14} \\
		\checkmark                          & \checkmark                           & \checkmark                              & \checkmark                                 &                &                             &                             & 91.85±0.26          & 93.19±0.56          & 93.62±0.15          & 93.87±0.26          \\
		\checkmark                           & \checkmark                           & \checkmark                             & \checkmark                                  & \checkmark             &                             &                             & 91.76±0.45          & 93.14±0.20          & 93.39±0.43          & 93.53±0.23          \\
		\checkmark                          & \checkmark                           & \checkmark                              &                                     &                &                             &                             & \textbf{92.16±0.33} & 93.18±0.11          & \textbf{93.83±0.17} & 93.82±0.11          \\ \hline
	\end{tabular}}
\end{table}

\subsubsection{Effect of the function used to calculate perturbation error} We propose the PCEM to quantify the impact of perturbations on the consistency of the model predictions. When evaluating the error introduced by a perturbation, we measured the dispersion between predicted and Paverage using the MSE loss function. This involves squaring the difference between each predicted value and the average value to magnify the corresponding error. However, there are several mainstream loss functions that can be used to measure dispersion, including Kullback-Leibler (KL) divergence and Huber Loss, etc. Each has its own advantages. To identify the most appropriate function for this framework, we compared several commonly used loss functions. The results in Table \RNum{5} demonstrate that our framework achieves higher accuracy compared to that randomly select data, given an equal number of annotated data, when using the loss functions presented in the table. The method based on MSE Loss maximizes our algorithm's performance. L1 Loss is an absolute value function, and compared to L1 Loss, MSE takes the square of the difference between the average and predicted values, which further amplifies the superiority of our approach, as indicated by the experimental results.

\begin{table}[h]
	\centering
	\caption{Abulation study on the function used to calculate perturbation error.}
	\scalebox{0.75}{
	\begin{tabular}{c|c|ccccc}
		\hline
		Method    & Metric & 10\%       & 20\%       & 30\%       & 40\%       & 50\%       \\ \hline
		Hinger Loss    &\multirow{6}{*}{Acc$\uparrow$}     & \multirow{6}{*}{89.40±0.94} & 91.06±0.41 & 91.89±0.14 & 92.49±0.59 & 92.52±0.61 \\
		KL divergence        &        &  & 91.97±0.34 & 92.97±0.47 & 93.21±0.55 & 93.57±0.22 \\
		L1 Loss       &        &  & 91.69±0.88 & 92.98±0.41 & 93.46±0.36 & 93.39±0.65 \\
		Smooth L1 Loss &        &  & 91.54±0.63 & 92.72±0.16 & 93.56±0.27 & 93.71±0.25 \\
		Huber Loss     &        &  & 91.89±0.66 & 93.02±0.46 & 93.33±0.21 & \textbf{93.84±0.29} \\
		MSE Loss       &        &  & \textbf{92.10±0.35} & \textbf{93.06±0.09} & \textbf{93.76±0.20} & 93.76±0.12 \\ \hline
		Hinger Loss    &\multirow{6}{*}{Pre$\uparrow$}    & \multirow{6}{*}{89.65±0.78} & 91.13±0.42 & 91.98±0.10 & 92.56±0.58 & 92.64±0.56 \\
		KL divergence        &        &  & 92.06±0.31 & 93.03±0.48 & 93.30±0.52 & 93.63±0.24 \\
		L1 Loss        &        &  & 91.09±0.71 & 93.01±0.42 & 93.51±0.34 & 93.44±0.63 \\
		Smooth L1 Loss &        &  & 91.68±0.52 & 92.79±0.19 & 93.61±0.26 & 93.77±0.25 \\
		Huber Loss     &        &  & 92.02±0.67 & 93.06±0.45 & 93.40±0.23 & \textbf{93.88±0.29} \\
		MSE Loss       &        &  & \textbf{92.16±0.33} & \textbf{93.18±0.11} & \textbf{93.83±0.17} & 93.82±0.11 \\ \hline
	\end{tabular}}
\end{table}
\section{Conclusion}
Given the diverse nature of medical imaging data derived from various devices and involving tasks with different 2D and 3D modalities, along with the need for algorithm generalization, we propose a straightforward AL-based method based on perturbation consistency. We believe that assigning a label to those with higher influence from perturbations holds greater value for existing annotated datasets, as it efficiently enhances data diversity and reduces annotation costs. Conversely, annotating datasets with lower perturbation influence provides limited benefits to the existing dataset and can be considered as an inefficient utilization of annotation resources. Experimental results on three different public datasets demonstrate the competitiveness and versatility of our proposed PCDAL framework. It can be applied to both 2D medical image classification and segmentation tasks, as well as 3D medical image segmentation tasks.
The framework reduced the amount of labeled data while improving algorithm performance, which was essential for developing computer-aided diagnosis systems.

Labeling datasets was crucial throughout the entire process, as sufficient data labeling was necessary for deep learning algorithms to realize their full performance potential. The PCDAL framework had enormous potential to provide guidance and important clues for constructing datasets in future computer-aided diagnosis systems that focused on various modalities of medical images and multiple recognition tasks based on 2D and 3D datasets. It could efficiently utilize doctors' precious time, reduce the cost of labeling, and decrease their workload for data labeling. The active learning method had already been applied to the dataset labeling process of the intelligent gastric cancer diagnosis platform at the Fujian Provincial Cancer Hospital in China.



%
%
%
%
%

%
%
\bibliography{references}
\end{document}